\documentstyle[12pt]{article}
\newcommand{\bea}{\begin{eqnarray}}
\newcommand{\eea}{\end{eqnarray}}
\newcommand{\be}{\begin{equation}}
\newcommand{\ee}{\end{equation}}
\newcommand{\p}{\prime}
\newcommand{\x}{\stackrel{\ss}{x}}
\newcommand{\q}{\stackrel{\ss}{q}}
\newcommand{\r}{\stackrel{\ss}{p}}
\newcommand{\nn}{\nonumber}
\newcommand{\rf}[1]{(\ref{#1})}

\begin{document}

\begin{center}
{\large\bf THE OSTROGRADSKY METHOD FOR LOCAL SYMMETRIES. CONSTRAINED
THEORIES WITH HIGHER DERIVATIVES}\\
\vspace*{1cm}
{\large N.P.Chitaia, S.A.Gogilidze}\\
{\it Tbilisi State University, University St.9, 380086 Tbilisi, Georgia,}\\
and \\
{\large{Yu.S.Surovtsev}} \\
{\it Joint Institute for Nuclear Research, Dubna 141 980, Moscow Region,
Russia}
\begin{abstract}
In the generalized Hamiltonian formalism by Dirac, the method of constructing
the generator of local-symmetry transformations for systems with first- and
second-class constraints (without restrictions on the algebra of constraints)
is obtained from the requirement for them to map the solutions of the
Hamiltonian equations of motion into the solutions of the same equations. It
is proved that second-class constraints do not contribute to the
transformation law of the local symmetry entirely stipulated by all the
first-class constraints (and only by them). A mechanism of occurrence of
higher derivatives of coordinates and group parameters in the symmetry
transformation law in the Noether second theorem is elucidated. It is shown
that the obtained transformations of symmetry are canonical in the extended
(by Ostrogradsky) phase space. An application of the method in theories with
higher derivatives is demonstrated with an example of the spinor Christ -- Lee
model.
\end{abstract}
\end{center}
{\bf 1.} When realizing various quantization schemes of gauge theories
(covariant, canonical, BRST), the explicit form of gauge transformations is
needed. At present there appear constrained theories for which the explicit
form of gauge transformations is unknown though the presence of first-class
constraints in a theory testifies to their existence. These are
theories considering, for example, a relativistic particle with torsion and
curvature, supersymmetric models, Siegel's symmetries, supergravity.
Therefore, it is desirable to have a method of constructing gauge
transformations on the given Lagrangian.

Earlier, for constrained special-form theories with first- and second-class
constraints (when the first-class primary constraints are the ideal of a
quasi-algebra of all the first-class constraints), we have suggested the
method of constructing local-symmetry transformations \cite{CGS-2}. However,
in the existing literature there are examples of Lagrangians where this
condition on constraints does not hold (see, e.g., \cite{CGS-3} and references
there). Then it was natural to ask: Can the local-symmetry transformations be
obtained in these theories? What is a role of second-class constraints under
these transformations and, generally, what is the nature of the Lagrangian
degeneracy in this case? For example, in ref.\cite{Gracia-Pons} it is stated
that in the mentioned example the gauge transformation generators do not exist
for the Hamiltonian formalism though for the Lagrangian one the gauge
transformations may be constructed; and in ref.\cite{Sugano-Kimura} one even
asserts that second-class constraints contribute also to a generator of gauge
transformations.

Here, suggesting the method of constructing local-symmetry transformations
in the general case without restrictions on their algebra, we answer the
formulated questions, manifest the mechanism of appearance of higher
derivatives of coordinates and group parameters in the transformation law and
show the obtained transformations to be canonical in the extended (by
Ostrogradsky) phase space.

{\bf 2.} In ref.\cite{CGS-1} we have suggested the method of the separation of
constraints into the first- and second-class ones by passing to the canonical
set of constraints, which is always possible. Therefore, below we consider a
dynamical system with the canonical set
$(\Phi_\alpha^{m_\alpha},\Psi_{a_i}^{m_{a_i}})$ of first- and second-class
constraints, respectively $(\alpha=1,\cdots,F,~ m_\alpha=1,\cdots,M_\alpha;~~
a_i=1,\cdots,A_i,~m_{a_i}=1,\cdots,M_{a_i},~i=1,\cdots,n)$, properties of
which are expressed by the following Poisson brackets:
\bea
&& \bigl\{\Phi_\alpha^{m_\alpha},H\bigr\} =
g_{\alpha~~\beta}^{m_\alpha m_\beta}~\Phi_\beta^{m_\beta},\quad
m_\beta=1,\cdots,m_\alpha+1,~~~~~~~~~~~~~~~~~~~ \label{PB-Phi-H^prime}\\
&& \bigl\{\Psi_{a_i}^{m_{a_i}},H\bigr\} = \bar{g}_{{a_i}~~\alpha}^{m_{a_i}
m_\alpha}~\Phi_\alpha^{m_\alpha}+\sum_{k=1}^{n}h_{{a_i}~~{b_k}}^{m_{a_i}
m_{b_k}}~\Psi_{b_k}^{m_{b_k}},\quad m_{b_n}=m_{a_i}+1,~~~~~~~\label{PB-Psi-H^prime}\\
&& \bigl\{\Phi_\alpha^{m_\alpha},\Phi_\beta^{m_\beta}\bigr\} =
f_{\alpha~~\beta~~\gamma}^{m_\alpha m_\beta m_\gamma}~\Phi_\gamma^{m_\gamma},
~~~~~~~~~~~~~~~~~~~\label{PB-Phi-Phi}\\
&& \bigl\{\Phi_\alpha^{m_\alpha},\Psi_{a_i}^{m_{a_i}}\bigr\}~
\stackrel{\Sigma}{=}0   \label{PB-Phi-Psi}
\eea
with
$$H = H_c+\sum_{k=1}^{n}({\bf K}^{1~k})_{b_k~a_k}^{-1}
\{\Psi_{a_k}^k,H_c\} \Psi_{b_k}^1$$
being a first-class function \cite{Dirac}; $H_c$ is the canonical Hamiltonian.
The general properties of the structure functions are given in work
\cite{CGS-1}. The canonical set is characterized by that each second-class
constraint has the vanishing Poisson brackets (on the constraint surface) with
all the constraints of the system except one, and the first-class constraints have the
vanishing Poisson brackets with all the constraints.

A group of phase-space coordinate transformations that maps each solution of
the Hamilton equations of motion into the solution of the same equations
will be called the symmetry transformation.

Consider the Hamilton equations of motion in the following form:
\bea \label{mot.eq-ns}
\left\{\begin{array}{l}
\dot q_i~\stackrel{\Sigma_1}{\approx}~\{q_i,H_T\},\quad \dot p_i ~\stackrel
{\Sigma_1}{\approx}~ \{p_i,H_T\}, \quad i=1,\cdots,N,\\ 
\Psi_{a_k}^1~\stackrel{\Sigma_1}{\approx}0,~~a_k=1,\cdots,A_k~(k=1,\cdots,n),\\
\Phi_\alpha^1~\stackrel{\Sigma_1}{\approx}0,~~ \alpha=1,\cdots,F,
\end{array}\right.
\eea
where $~~H_T = H + u_\alpha\Phi_\alpha^1;~$
$u_\alpha$ are undetermined Lagrange multipliers; the symbol
$\stackrel{\Sigma_1} {\approx}$ means weak equality on the primary-constraints
surface $\Sigma_1$.

Consider also the transformations of the phase-space coordinates
\bea \label{q-q^prime}
\left\{\begin{array}{ll} q_i^\p = q_i+\delta q_i ,& \quad
\delta q_i = \{q_i , G\} ,
\\ p_i^\p = p_i+\delta p_i ,& \quad\delta p_i =
\{p_i , G\} \end{array}\right.
\eea
with the generator $G$ sought in the form
\be \label{G-Dirac}
G=\varepsilon_\alpha^{m_\alpha}\Phi_\alpha^{m_\alpha} +
\eta_{a_i}^{m_{a_i}}\Psi_{a_i}^{m_{a_i}}.
\ee
To recognize a role of the second-class constraints in gauge transformations
in this general case, we consider them on the same basis as the first-class
constraints.

We will require the transformed quantities $q_i^\p(t)$ and $p_i^\p(t)$ defined
by \rf{q-q^prime} to be solutions of the Hamilton equations of motion
\rf{mot.eq-ns} provided that the initial $q_i(t)$ and $p_i(t)$ do this, i.e.
\bea \label{mot.eq-ns'prime}
&&\dot q_i^\p~\stackrel{\Sigma_1}{\approx}~\frac{\partial H_T^\p}
{\partial p_i}(q^\p,p^\p),\qquad \dot p_i^\p ~\stackrel {\Sigma_1}{\approx}~
-\frac{\partial H_T^\p}{\partial q_i}(q^\p,p^\p), \quad i=1,\cdots,N,\\
&&\Psi_{a_k}^1(q^\p,p^\p)~\stackrel{\Sigma_1}{\approx}0,~~a_k=1,\cdots,A_k~
(k=1,\cdots,n),\\
&&\Phi_\alpha^1(q^\p,p^\p)~\stackrel{\Sigma_1}{\approx}0,~~ \alpha=1,\cdots,F,
\eea
where
$~~~H_T^\p=H_T+\delta u_\alpha(t)\Phi_\alpha^1(q,p)=H+
u_\alpha^\p(t)\Phi_\alpha^1 (q,p).$~ Here it is taken into consideration that
the transformations \rf{q-q^prime} must conserve the primary-constraints
surface $\Sigma_1$. This can easily be interpreted if one remembers that
$\Sigma_1$ is the whole $(q,\dot q)$-space image in the phase space.
Since under the operation of the local-symmetry transformation group the
$(q,\dot q)$-space is mapped into itself in a one-to-one manner, therefore,
the one-to-one mapping of $\Sigma_1$ into itself corresponds to that in the
phase space. Therefore, it is natural to require
\be \label{PB-G-Psi1,Phi1}
\bigl\{\Psi_{a_k}^1,G\bigr\}~\stackrel{\Sigma_1}{=}0,\qquad
\bigl\{\Phi_\alpha^1,G\bigr\}~\stackrel{\Sigma_1}{=}0.
\ee
The realization of the first condition \rf{PB-G-Psi1,Phi1} would mean
\cite{CGS-3}
\be \label{eta-i=0}
\eta_{a_i}^i=0\qquad \mbox{for}~~ i=1,\cdots,n.
\ee
As to the second condition \rf{PB-G-Psi1,Phi1}, we rewrite it in the form:
\be \label{PB-G-Phi1-detail}
\bigl\{\Phi_\alpha^1,G\bigr\}=\bigl(f_{\alpha~~\beta~~\gamma}^{1~ m_\beta~1}~
\Phi_\gamma^1+f_{\alpha~~\beta~~\gamma}^{1~~m_\beta~m_\gamma}~\Phi_\gamma^
{m_\gamma}\bigr)\varepsilon_\beta^{m_\beta}+\bigl\{\Phi_\alpha^1,\Psi_{a_i}^
{m_{a_i}}\bigr\}\eta_{a_i}^{m_{a_i}}~\stackrel{\Sigma_1}{=}0,~
\ee
The last term in \rf{PB-G-Phi1-detail} vanishes due to \rf{PB-Phi-Psi};
therefore, the equality \rf{PB-G-Phi1-detail} would be satisfied if
~$f_{\alpha~~\beta~~\gamma}^{1~~m_\beta~m_\gamma}=0$~ for $m_\gamma\geq 2$.
This case is considered in \cite{CGS-2}. Here we consider the general case
when
$$f_{\alpha~~\beta~~\gamma}^{1~~m_\beta~m_\gamma}\neq 0
\quad\mbox{for}\quad m_\gamma \geq 2.$$
One can show \cite{CGS-3} that one can always pass to an equivalent set of
constraints $~\tilde{\Phi}_\beta^{m_\beta}~$ for which
~$~\tilde{f}_{\alpha~~\beta~~\gamma}^{1~~m_\beta~m_\gamma}=0$~~ (for
$m_\gamma\geq 2$) and
\be \label{ideal}
\bigl\{\tilde{\Phi}_\alpha^1,\tilde{\Phi}_\beta^{m_\beta}\bigr\}=
\tilde{f}_{\alpha~~\beta~~\gamma}^{1~~m_\delta~1}~\tilde{\Phi}_\gamma^1,
\ee
which is needed for the realization of \rf{PB-G-Phi1-detail}. The
corresponding transformation is
$$
\tilde{\Phi}_\beta^{m_\beta}=C_{\beta~~\alpha}^{m_\beta m_\alpha}\Phi_\alpha^
{m_\alpha},\qquad \mbox{det}\left\|C_{\beta~~\alpha}^{m_\beta
m_\alpha}\right\|_\Sigma \not =0,$$
where
$~C_{\beta~~\alpha}^{1~~m_\alpha}=\delta_{\beta\alpha}~\mbox{for any}
~m_\alpha;~$
and elements $~C_{\beta~~\gamma}^{m_\beta m_\gamma}~$ for
~$m_\beta,m_\gamma\geq 2~$ satisfy the system of equations
$$\bigl\{\Phi_\alpha^1,C_{\beta~~\gamma}^{m_\beta m_\gamma}\bigr\}+
f_{\alpha~~ \delta~~\gamma}^{1~~m_\delta m_\gamma}C_{\beta~~\delta}^
{m_\beta m_\delta}=0,$$
which is fully integrable \cite{CGS-3}. Considering this transition to be
carried out, we shall below omit the mark ``$\>\tilde{\;}\>$''.

Now from \rf{mot.eq-ns},\rf{mot.eq-ns'prime} and the Jacobi identities
for $(q_i,G,H_T^\p)$ and $(p_i,G,H_T^\p)$ and taking into account
\rf{PB-Phi-H^prime} - \rf{PB-Phi-Psi} and \rf{PB-G-Psi1,Phi1}, we obtain
\cite{CGS-3}:
\be \label{eps.eta-Phi-Psi}
\Bigl (\dot\varepsilon_\alpha^{m_\alpha}+\varepsilon_\beta^{m_\beta}
g_{\beta~~\alpha}^{m_\beta m_\alpha}+\sum_{i=1}^{n}\eta_{a_i}^{m_{a_i}}
\bar{g}_{{a_i}~~\alpha}^{m_{a_i} m_\alpha}~\Bigr)\Phi_\alpha^{m_\alpha}
+\sum_{i,k=1}^{n}\Bigl(\dot{\eta}_{a_i}^{m_{a_i}}+\eta_{b_k}^{m_{b_k}}
h_{{b_k}~~{a_i}}^{m_{b_k} m_{a_i}}~\Bigr)\Psi_{a_i}^{m_{a_i}}
~\stackrel{\Sigma_1}{=}0.
\ee
In view of the functional independence of constraints $\Phi_\alpha^{m_\alpha}$
and $\Psi_{a_i}^{m_{a_i}}$, in order to satisfy the equality
\rf{eps.eta-Phi-Psi}, one must demand the coefficients of constraints
$\Phi_\alpha^{m_\alpha}~(m_\alpha \geq 2)$ and $\Psi_{a_i}^{m_{a_i}}~(m_{a_i}
\geq 2)$ to vanish. Consider the requirement of vanishing the coefficients of
constraints $\Psi_{a_i}^{m_{a_i}}~(m_{a_i} \geq 2,~i=2,\cdots,n,~a_i=
1,\cdots,A_i)$:
\bea \label{eq:eta}
\dot{\eta}_{a_i}^{m_{a_i}}+\sum_{k=1}^{n}\eta_{b_k}^{m_{b_k}}
h_{{b_k}~~{a_i}}^{m_{b_k} m_{a_i}}=0.
\eea
Taking \rf{eta-i=0} into account, one can see that this system of equations
has only a trivial solution, since one can show by the proof by contradiction
\cite{CGS-2} that
$$~~\mbox{det}\|h_{b_i~~~~~a_i}^{i-k-1~i-k}\|\neq 0.$$
Therefore, all quantities $\eta_{a_i}^{m_{a_i}}$ vanish, i.e. the
second-class constraints do not contribute to the generator of local-symmetry
transformations unlike the assertions in refs.\cite{Sugano-Kimura}.

So, to satisfy the equality \rf{eps.eta-Phi-Psi}, we must now require
vanishing the coefficients of $\Phi_\alpha^{m_\alpha}$ in
\rf{eps.eta-Phi-Psi}, i.e. determine the multipliers
$\varepsilon_\alpha^{m_\alpha}$ in the generator \rf{G-Dirac} from the system
of equations
\be \label{eq:eps}
\dot \varepsilon_\alpha^{m_\alpha}+\varepsilon_\beta^{m_\beta}
g_{\beta~~\alpha}^{m_\beta m_\alpha}=0,\qquad m_\beta= m_\alpha-1,\cdots,
M_\alpha,
\ee
which is solved by the procedure of reparametrization with introducing
$F$ arbitrary functions $\varepsilon_\alpha(t)\equiv \varepsilon_\alpha^
{M_\alpha}$ \cite{GSST:tmf1}. As a result, we obtain the generator in the form
\be \label{G}
G=B_{\alpha~~\beta}^{m_\alpha m_\beta}\phi_\alpha^{m_\alpha}\varepsilon_\beta^
{(M_\alpha-m_\beta)} , \qquad m_\beta=m_\alpha,\cdots,M_\alpha ,
\ee
where
$~\varepsilon_\beta^{(M_\alpha-m_\beta)}\equiv
(d^{M_\alpha-m_\beta}/dt ^{M_\alpha-m_\beta})\varepsilon_\beta
(t);~$ and $~B_{\alpha~~\beta}^{m_\alpha m_\beta}~$ are, generally speaking,
functions of $q$ and $p$ and their derivatives up to the order
$M_\alpha-m_\alpha-1$.

The corresponding transformations of local symmetry in the Lagrangian
formalism are determined in the following way
$$\delta q_i(t)=\{q_i(t),G\}\biggr|_{p=\partial L/
\partial \dot q}, \qquad \delta\dot q(t) = \frac{d}{dt}\delta q(t). $$

When deriving the gauge transformation generator the employment of the
obtained system \rf{eq:eps} is important, the solution of which manifests the
mechanism of appearance of higher derivatives of coordinates and group
parameters in the Noether transformation law in the configuration space,
the highest possible order of coordinate derivatives being determined by the
structure of the first-class constraint algebra, and the order of the highest
derivative of group parameters in the transformation law equals
($M_\alpha-1$).

{\bf 3.} When a generator $G$ depends on higher derivatives of coordinates and
momenta (e.g., Polyakov's string and others \cite{CGS-3}),
this indicates that gauge transformations in the configuration
space also depend on higher derivatives of $q$. Under those transformations
we have
\be \label{L-prime}
L^\prime =L(q ,\dot q) + {d\over dt}F(q ,\dot q,\ddot q ,\cdots, \varepsilon,
\dot \varepsilon ,\cdots)
\ee
where $\varepsilon(t)$ are the group parameters. Adding to Lagrangian
$L(q,\dot q)$ the total time derivative of a function which depends also on
higher derivatives does not change the Lagrange equations of motion. But
$L^\p$ is the Lagrangian with higher derivatives. Therefore, the Hamiltonian
formulation of the theory with the Lagrangian $L$ must be built in the same
extended (by Ostrogradsky) phase space as it is the case for $L^\p$.

We construct the extended phase space using the formalism of theories with
higher derivatives \cite{Ostr} and determine the coordinates as follows
\be \label{enlarged-q}
q_{1~i}=q_i ,\quad q_{s~i}={d^{s-1}\over{dt^{s-1}}}~q_i ,\quad s=2 ,\cdots, K ,
\quad i=1 ,\cdots, N
\ee
where $K$ equals the highest order of derivatives of $q$ and $p$. The
conjugate momenta defined by the formula of theories with higher derivatives
\cite{Ostr}
\be \label{Ostr.:p}
p_{r~i}=\sum_{l=r}^{K}(-1)^{l-r}\frac{d^{l-r}}{dt^{l-r}}\frac{\partial L}
{\partial q_{r+1~i}}
\ee
are
$$p_{1~i}=p_i ,\qquad p_{s~i}=0\quad\mbox{ for }\quad s=2,\cdots, K.$$
The generalized momenta for $s\geq2$ are extra primary constraints of the
first class. Now the total Hamiltonian is written down as
$$\overline{H}_T = H_T(q_{1~i} , p_{1~i}) +
\lambda_{s~i}~p_{s~i}, \quad s\geq 2,$$
where $H_T$ is of the same form as in the initial phase space and
$\lambda_{s~i}$ are arbitrary functions of time. Additional secondary
constraints corresponding to $p_{s~i}$ for $s\geq 2$ do not appear.

Now the Poisson brackets are determined in the following way
$$\bigl\{A,B\bigr\}=\frac{\partial A} {\partial q_{r~i}}\frac{\partial B}
{\partial p_{r~i}}-\frac{\partial A}{\partial p_{r~i}}\frac{\partial B}
{\partial q_{r~i}}.$$

Further, constructing gauge transformations, one must operate by
the described method. However, before to determine
$\varepsilon_\alpha^{m_\alpha}$ from the system of equations \rf{eq:eps},
for derivatives of $p_{1~i}$ we shall make the following replacements:
\bea \label{dot-p}
p_{1~i} & = & \frac{\partial L}{\partial q_{2~i}} = h_0^i (q_{1~k} , q_{2~k}) ,
\qquad i,k=1,\cdots,N,\nn\\
\dot p_{1~i} & = & \frac{\partial h_0^i}{\partial q_{1~n}} q_{2~n} +
\frac{\partial h_0^i}{\partial q_{2~n}} q_{3~n} = h_1^i(q_{1~k} , q_{2~k} ,
q_{3~k}) ,\\ & \vdots & \nn\\
p_{1~i}^{(M_\alpha - 2)} & = & h_{M_\alpha - 2}^i (q_{1~k} , q_{2~k} ,\cdots,
q_{M_\alpha-1~k}) .\nn
\eea
So, as a result, we shall obtain the generator $\overline{G}$ in the extended
phase space
\be \label{enl.G}
\overline{G}=B_{\alpha~~\beta}^{m_\alpha m_\beta}\Phi_\alpha^{m_\alpha}
\varepsilon_\beta^{(M_\alpha-m_\beta)} + \varepsilon_{s~i}p_{s~i},~~~
m_\beta=m_\alpha,\cdots,M_\alpha,~~  s=2 ,\cdots, K,
\ee
where $B_{\alpha~~\beta}^{m_\alpha m_\beta}(q_{1~i},\cdots,q_{M_\alpha -1~i};
p_{1~i})$, being just in the same forms as in the initial phase space, are
written, however, with taking account of the above-indicated replacements;
$\varepsilon_{s~i}$ are supplementary group parameters. Note that the
obtained generator \rf{enl.G} satisfies the group
property ~$\{\overline{G}_1,\overline{G}_2\}=\overline{G}_3.$~
Now the local-symmetry transformations of the coordinates of the
initial phase space in the extended one are of the form
\bea
\label{enl.-delta-q-p}
\left\{\begin{array}{l}
\delta q_{1~k}=\varepsilon_\beta^{(M_\alpha-m_\beta)}
\bigl\{q_{1~k},B_{\alpha~~\beta}^{m_\alpha m_\beta}(q_{1~i},\cdots,
q_{M_\alpha -1~i} ; p_{1~i})\phi_\alpha^{m_\alpha}(q_{1~i},p_{1~i})\bigr\},\\
{}\\
\delta p_{1~k}=\varepsilon_\beta^{(M_\alpha-m_\beta)}\bigl\{p_{1~k},
B_{\alpha~~\beta}^{m_\alpha m_\beta}(q_{1~i},\cdots,q_{M_\alpha -1~i} ;
p_{1~i})\phi_\alpha^{m_\alpha}(q_{1~i},p_{1~i})\bigr\}.
\end{array}\right.
\eea
To within quadratic terms in $\delta q_{i~k}$ and $\delta p_{j~n},$
$$\{q_{i~k}+\delta q_{i~k} ,p_{j~n}+\delta p_{j~n}\}=\delta_{ij}\delta_{kn},$$
{\it i.e.}, the obtained transformations of local symmetry are canonical in
the extended (by Ostrogradsky) phase space.

{\bf 4. Example.} This method of constructing the generator of gauge
transformations can also be applied in theories with higher derivatives,
where in constructing the phase space one must use the Ostrogradsky method.
We demonstrate this with an example of the spinor Christ -- Lee model.
Consider the Lagrangian
\be \label{Christ-Lee:L}
L=\frac{1}{2}\Bigl[\bigl(\partial_t+yT\bigr)^2\x\Bigr]^2+
\psi^+\bigl(i\partial_t+yç\bigr)\psi+m\psi^+\psi,
\ee
where
$$ \x=\left(\begin{array}{c}x_1\\x_2 \end{array}\right),\quad
T=\left(\begin{array}{cr}
0 & -1\\1 & 0 \end{array}\right),\quad
ç=\left(\begin{array}{cr}
1 & 0\\0 & -1 \end{array}\right).$$
Since $L$ in \rf{Christ-Lee:L} depends on $\ddot{\x}$, passing to the
Hamiltonian formalism we construct the extended phase space (by Ostrogradsky)
as follows: We determine the coordinates
$$\q_1=\x, \quad \q_2=\dot{\x}, \quad q_1=y, \quad q_2=\dot y, \quad \psi
\quad \mbox{and} \quad \psi^+.$$
Then the generalized momenta are determined by formula \rf{Ostr.:p};
and we obtain three primary constraints
$$\phi_1^1=p_2,\quad \phi_2^1=\pi_\psi-i\psi^+,\quad \phi_3^1=\pi_{\psi^+}$$
and
$$H_c = \frac{1}{2} \r_2^2 + \r_1\q_2 + p_1q_2 - q_1^2\r_2\q_1 -
2q_1 \r_2T\q_2 - q_2\r_2T\q_1 - q_1\psi^+ç\psi - m\psi^+\psi.$$
Among the conditions of the constraint conservation in time
~$\dot{\psi}_i^1= 0~(i=1,2,3)$~ the last two ones serve for determining the
Lagrange multipliers. From the first condition we obtain one secondary
constraint $$\phi_1^2=\r_2T\q_1-p_1$$ and the tertiary constraint
$$\phi_1^3=-\r_2T\q_2-\r_1T\q_1-i\pi_\psi \psi $$
and there do not arise more constraints. Calculating the matrix ~${\bf W}=
\left\|\{\phi_\alpha^{m_\alpha},\phi_\beta^{m_\beta}\}\right\|$, we obtain
that ~$\mbox{rank}{\bf W}=2$; therefore, two constraints are of second class
and the three ones are of first class. Now we pass to the canonical set of
constraints by the equivalence transformation:
\bea
\left(\begin{array}{c}
\Psi_1^1\\ \Psi_2^1\\ \Phi_1^1\\ \Phi_1^2 \\ \Phi_1^3\end{array}\right)
&=& \left(\begin{array}{ccccc}
1 & 0 & 0 & ~~0 & ~~0 \\
0 & 1 & 0 & ~~0 & ~~0 \\
0 & 0 & 1 & ~~0 & ~~0 \\
0 & 0 & 0 & ~~1 & ~~0 \\
i\psi & -i\psi^+ & 0 & ~~0 & ~~1
\end{array}  \right)
\left(\begin{array}{c}
\phi_2^1\\ \phi_3^1\\ \phi_1^1\\ \phi_1^2 \\ \phi_1^3\end{array}  \right)=\nn\\
&=&\left(\begin{array}{c}
\pi_\psi-i\psi^+\\ \pi_{\psi^+} \\ p_2\\ \r_2T\q_1-p_1\\
-\r_2T\q_2-\r_1T\q_1-i(\pi_\psi \psi+\psi^+\pi_{\psi^+})
\end{array}  \right),
\eea
where the constraints are already separated into the ones of first and second
class.

Further, we seek the generator $G$ in the form
$$G=\eta_1^1~\Psi_1^1+\eta_2^1~\Psi_2^1+\varepsilon_1^{m_1}~\Phi_1^{m_1},
~~m_1=1,2,3.$$
>From the second condition \rf{PB-G-Psi1,Phi1} of the $\Sigma_1$ conservation
we derive ~$\eta_1^1=\eta_2^1=0,$  {\it i.e.}, the constraints of second class
do not contribute to $G$. The first condition \rf{PB-G-Psi1,Phi1} is realized
because
$$\bigl\{\Phi_1^1,\Phi_1^{m_1^\p}\bigr\}=0 ~~(m_1^\p=2,3).$$
Taking into account that ~$g_{1~1}^{1~2}=g_{1~1}^{2~3}=1$~ and all other
$g$'s vanish in \rf{PB-Phi-H^prime}, eq.\rf{eq:eps} accepts the form:
$$\dot \varepsilon_1^2 -\varepsilon_1^1 = 0, \qquad \dot \varepsilon_1^3
-\varepsilon_1^2 = 0,$$ {\it i.e.}, ~$\varepsilon_1^2=\dot \lambda$~ and
~$\varepsilon_1^1=\ddot\lambda$~ where ~$\lambda\equiv \varepsilon_1^3$.~
Therefore we have $$G=\ddot\lambda(t) p_2+\dot\lambda(t)\bigl(r_2T\q_1-p_1
\bigr)-\ddot\lambda(t)\bigl[\r_2T\q_2+\r_1T\q_1+i(\pi_\psi\psi+\psi^+
\pi_{\psi^+})\bigr],$$ from which it is easily to obtain the gauge
transformations in the phase space and the following transformations in the
configuration space
$$\delta y=\dot\lambda(t),\quad \delta\x=\lambda(t)T\x, \quad
\delta \psi=i\lambda(t)\psi, \quad \delta\psi^+=-i\lambda(t)\psi^+$$
and to verify that the action is invariant under these transformations
($\delta L=0$).

{\bf 5.} So, we can state in the general case of theories with first- and
second-class constraints (without restrictions on the constraint algebra) that
the necessary and sufficient condition for a certain quantity $G$ to be a
local-symmetry transformation generator is the representation of $G$ as a
linear combination of all the first-class constraints (and only of them) of
the equivalent set of the special form (when the condition \rf{ideal} is
valid) with the coefficients determined by the system of equations
\rf{eq:eps}. (Passing to the indicated equivalent set of constraints is always
possible, and the method is presented in \cite{CGS-3}.) In addition, these are
the necessary and sufficient conditions for \rf{q-q^prime} to be the
quasi-invariance transformation of the action functional in both the phase and
$(q,\dot q)$ space. It is thereby shown in the general case that the action
functional and the corresponding Hamiltonian equations of motion are invariant
under the same quasigroup of local-symmetry transformations, and the
degeneracy of theories with the first- and second-class constraints is due to
their invariance under local-symmetry transformations.

One of the authors (S.A.G.) thanks the Russian Foundation for Fundamental
Research (Grant N$^{\underline {\circ}}$ 96-01-01223) for support.

\end{document}